\documentclass{elsart3}

\newcommand{\Journal}[4]{#1 #2 (#4) #3}
\newcommand{\PRev}{Phys. Rev.}
\newcommand{\JPSJ}{J. Phys. Soc. Jpn.}
\newcommand{\be}{\begin{equation}}
\newcommand{\ee}{\end{equation}}
\newcommand{\nn}{\nonumber}
\newcommand{\mib}[1]{\mathbf{#1}}
\newcommand{\cd}{Cd$_2$Re$_2$O$_7$}

\usepackage{amssymb}

\begin{document}

\begin{frontmatter}



\title{Structural order parameter and itinerant electron magnetism in \cd}

\author{I. A. Sergienko}, 
\ead{sergienko@physics.mun.ca}
\author{S. H. Curnoe}
\address{Department of Physics and Physical
Oceanography, Memorial University of Newfoundland,\\ St. John's, NL, A1B 3X7,
Canada}%




\begin{abstract}
We describe the low temperature behaviour of the magnetic susceptibility 
of \cd\ in terms of a Landau theory of structural phase transitions.
We calculate the zero temperature Pauli susceptibility using a tight-binding 
approach to reveal the mechanism of coupling between the structural
order parameter and itinerant magnetism.
\end{abstract}

\end{frontmatter}

\cd\ is the only pyrochlore superconductor~\cite{hanawa,sakai,jin1}. 
Upon cooling, \cd\ undergoes two structural phase transitions (SPT's) 
well above the very low temperature superconducting phase transition~\cite{gaulin,yamaura}. 
The first occurs
at 200 K and  is of second-order, 
while the other is 
first-order at 120 K. 
Neither SPT involves multiplication of the bcc primitive cell. 
We will refer to the phases as I, II and
III,  with corresponding space groups
Fd$\overline3$m$\rightarrow$I$\overline4$m$2 \rightarrow $I$4_122$ respectively.
The SPT's are accompanied by anomalous behaviour of physical properties, 
including the electrical resistivity, magnetic susceptibility, 
Hall coefficient and thermoelectric power. In this paper we concentrate on
the magnetic properties of \cd.

\cd\ is a paramagnetic metal in the normal state.
Above 400 K, the magnetic susceptibility displays approximate
Curie-Weiss behaviour with a large, negative Weiss temperature~\cite{sakai1}. 
On cooling it  reaches a broad maximum, and then falls off
below 200 K~\cite{hanawa,sakai1,jin}. No anomaly has been found at
the second SPT at 120 K. The kind of anomaly observed at 200 K is often
associated with an antiferromagnetic phase transition, 
but Re nuclear quadrupole 
resonance experiments find no evidence of magnetic ordering  in the low 
temperature phases~\cite{vyas}, thus revealing the itinerant electronic
nature of its paramagnetism.
 
In this article we show theoretically that the magnetic 
anomaly at 200 K is due to
the electrostatic interaction between itinerant electrons 
and tetragonal distortions of the pyrochlore lattice. 
Recently, we found a single two-dimensional order 
parameter (OP) that describes both SPT's
in \cd\ which corresponds to an instability of the lattice with respect to
a long wavelength phonon mode of $E_u$ symmetry~\cite{we}. 


We introduce the two 
components of the structural OP $(\eta_1, \eta_2)$,  which
are linear combinations of the displacements of four 
Re atoms, corresponding to one bcc site, from their ideal positions in phase 
I, $(x_m, y_m, z_m), m=1,2,3,4$~\cite{we}.
\be
\label{eta_op}
\eta_1=(X-Y)/\sqrt{2},\quad \eta_2=(X+Y-2Z)/\sqrt{6},
\ee
where $X=(x_1+x_2-x_3-x_4)/2$, $Y=(y_1-y_2+y_3-y_4)/2$ and 
$Z=(z_1-z_2-z_3+z_4)/2$. Phase II is characterized by $\eta_1=0$, $\eta_2\ne0$ 
and phase III by 
$\eta_1\ne0$, $\eta_2=0$. The OP spans the $E_u$ representation 
of the cubic point group $O_h$.
The magnetisation $\mib{M}$ corresponds to the
irreducible representation $F_{1g}$. 
Considering the symmetric products
$[E_u\otimes E_u]=A_{1g}\oplus E_g$ and $[F_{1g}\otimes F_{1g}]=A_{1g}\oplus
E_g\oplus F_{2g}$, we find that there are two independent constants
describing biquadratic coupling between the OP and $\mib{M}$.
In an applied magnetic field $\mib{H}$, the magnetic part of the free energy is
\begin{eqnarray}
\label{energy}
F_M &=& A\mib{M}^2-\mib{MH} +\gamma (\eta_1^2+\eta_2^2)\mib{M}^2\nn\\
&&+ \delta [(\eta_1^2-\eta_2^2)(2M_z^2-M_x^2-M_y^2)\\ 
&&\qquad + 2\sqrt{3}\eta_1\eta_2(M_x^2-M_y^2)].\nn
\end{eqnarray}
In all three phases the product $\eta_1\eta_2$ is zero, and the inverse
magnetic 
susceptibility is calculated to be
\begin{eqnarray}
\label{sus}
\chi_{zz}^{-1}=\chi_0^{-1}+2\gamma(\eta_1^2+\eta_2^2)
+4\delta(\eta_1^2-\eta_2^2)\nn\\
\chi_{xx}^{-1}=\chi_0^{-1}+2\gamma(\eta_1^2+\eta_2^2)
-2\delta(\eta_1^2-\eta_2^2)
\end{eqnarray}
where $\chi_0=(2A)^{-1}$ is the susceptibility of the high symmetry phase. 
In order to account for the experimentally observed behaviour at the two SPT's,
we should assume $\delta\ll\gamma$ and $\gamma>0$ \cite{we}.
Note the similarity between the present theory and the antiferromagnetic case,
where the anomaly in the susceptibility also arises from biquadratic coupling
between $\mib{M}$ and the antiferromagnetic vector.  


The phenomenological model~(\ref{energy}) is based on group-theoretical 
arguments only and does not distinguish between localised and itinerant 
electronic magnetism. We now restrict our attention to the itinerant case.
The tight-binding approach describes the dispersion of 
electronic bands $\epsilon_\mib{k}$ resulting from the electrostatic 
potential of the lattice.
Overlap integrals, which constitute matrix elements of the tight-binding 
Hamiltonian, depend on the radius vector between neighbouring 
atoms~\cite{slater}. Due to the relations~(\ref{eta_op}), this dependence
can be expressed in terms of the OP, $\epsilon_\mib{k}(\eta_1,\eta_2)$.

For simplicity we consider a simple one-band model for small $\mib{k}$
around the center of the Brillouin zone. Group-theoretical considerations
yield
\begin{eqnarray}
\label{band}
\epsilon_\mib{k}(\eta_1,\eta_2) &=& \hbar^2/2m^*[\mib{k}^2+
\alpha(\eta_1^2+\eta_2^2)\mib{k}^2\nn\\
&& +\beta [(\eta_1^2-\eta_2^2)(2k_z^2-k_x^2-k_y^2) \nn\\
&& +2\sqrt{3}\eta_1\eta_2(k_x^2-k_y^2)].
\end{eqnarray}
Here $m^*$ is the 
effective mass in phase I, and $\alpha$ and $\beta$ are constants which may 
be calculated by diagonalising the tight-binding Hamiltonian. Thus 
the effective mass components 
depend on the OP. In all three phases 
$\eta_1\eta_2=0$, therefore $m_x=m_y\neq m_z$. 

The density of states at the Fermi level (including spin) is
\begin{eqnarray}
\label{den}
D(E_F)&=&\frac{1}{\hbar^2}\left(\frac{3m_x^2m_zN}{\pi^4V}\right)^{1/3}\nn\\
&\approx&\frac{m^*}{\hbar^2}\left(\frac{3N}{\pi^4V}\right)^{1/3}
[1-\alpha(\eta_1^2+\eta_2^2)],
\end{eqnarray}
where $N/V$ is the number of conduction electrons per unit volume.
Then the  Pauli susceptibility at zero temperature is~\cite{white}
\be
\chi_{P}=\frac{g^2\mu_B^2m^*}{4\hbar^2}\left(\frac{3N}{\pi^4V}\right)^{1/3}
[1-\alpha(\eta_1^2+\eta_2^2)].  \label{pauli}
\ee
We note that up to the second order in the OP, only the isotropic part of the
effective mass contributes to the density of states 
and Pauli susceptibility, but
higher order corrections 
will yield an anisotropic contribution from
the OP. Moreover, spin-orbit coupling,
according to band structure calculations~\cite{harima,singh},
has a significant effect on
the electronic spectrum, and
also gives rise to anisotropy of the $g$ factor~\cite{white}. 
Nevertheless, the 
simple model proposed here provides a mechanism of coupling between
itinerant electrons and structural distortions,
and the lowest order result (\ref{pauli}) supports our assumption
about the relative sizes of the phenomenological constants in~(\ref{sus}).

This work was supported by NSERC of Canada.

\end{document}